\documentclass[12pt,a4paper]{article}
\usepackage{amsmath}
\usepackage{latexsym}
\usepackage{amssymb}
\usepackage{graphicx,color}
\makeatletter
\def\rddots{\mathinner{\mkern1mu\raise\p@%
    \vbox{\kern7\p@\hbox{.}}\mkern2mu%
    \raise4\p@\hbox{.}\mkern2mu\raise7\p@\hbox{.}\mkern1mu}}
\makeatother
\makeatletter
\def\eqnarray{%
\stepcounter{equation}%
\let\@currentlabel=\theequation
\global\@eqnswtrue
\global\@eqcnt\z@
\tabskip\@centering
\let\\=\@eqncr
$$\halign to \displaywidth\bgroup\@eqnsel\hskip\@centering
$\displaystyle\tabskip\z@{##}$&\global\@eqcnt\@ne
\hfil$\displaystyle{{}##{}}$\hfil
&\global\@eqcnt\tw@$\displaystyle\tabskip\z@{##}$\hfil
\tabskip\@centering&\llap{##}\tabskip\z@\cr}
\makeatother
\newcommand{\ket}[1]{{\vert{#1}\rangle}}

\newcommand{\fukuso}{{\mathbf C}}

\begin{document}

\title{\sl General Solution of the Quantum Damped \\
Harmonic Oscillator}
\author{
  Ryusuke ENDO
  \thanks{E-mail address : endo@sci.kj.yamagata-u.ac.jp }\ \ ,\ \ 
  Kazuyuki FUJII
  \thanks{E-mail address : fujii@yokohama-cu.ac.jp }\quad and\ \ 
  Tatsuo SUZUKI
  \thanks{E-mail address : suzukita@gm.math.waseda.ac.jp }\\
  ${}^{*}$Department of Physics\\
  Yamagata University\\
  Yamagata, 990-8560\\
  Japan\\
  ${}^{\dagger}$Department of Mathematical Sciences\\
  Yokohama City University\\
  Yokohama, 236--0027\\
  Japan\\
  ${}^{\ddagger}$Center for Educational Assistance\\
  Shibaura Institute of Technology\\
  Saitama, 337--8570\\
  Japan\\
  }
\date{}
\maketitle
%
%
%
%
\begin{abstract}
  In this paper the general solution of the quantum damped harmonic 
  oscillator is given.
\end{abstract}
%


%
%
%
%

\vspace{10mm}
In this paper we revisit dynamics of a quantum open system. First of all 
we explain our purpose in a short manner. See \cite{BP} as a general 
introduction to this subject. 
We consider a quantum open system $S$ coupled to the environment $E$. 
Then the total system $S+E$ is described by the Hamiltonian
\[
H_{S+E}=H_{S}\otimes {\bf 1}_{E}+{\bf 1}_{S}\otimes H_{E}+H_{I}
\]
where $H_{S}$, $H_{E}$ are respectively the Hamiltonians of the system and 
environment, and $H_{I}$ is the Hamiltonian of the interaction.

\par \noindent
Then under several assumptions (see \cite{BP}) the reduced dynamics of the 
system (which is not unitary !) is given by the Master Equation
\begin{equation}
\label{eq:master-equation}
\frac{\partial}{\partial t}\rho=-i[H_{S},\rho]-{\cal D}(\rho)
\end{equation}
with the dissipator being the usual Lindblad form
\begin{equation}
\label{eq:dissipator}
{\cal D}(\rho)=\frac{1}{2}\sum_{\{j\}}
\left(A_{j}^{\dagger}A_{j}\rho+\rho A_{j}^{\dagger}A_{j}
-2A_{j}\rho A_{j}^{\dagger}\right).
\end{equation}
Here $\rho\equiv \rho(t)$ is the density operator (or matrix) of the system. 

Similarly, the equation of quantum damped harmonic oscillator (see \cite{BP}, 
Section 3.4.6) is given by
\begin{equation}
\label{eq:quantum damped harmonic oscillator}
\frac{\partial}{\partial t}\rho=-i[\omega a^{\dagger}a,\rho]
-
\frac{\mu}{2}
\left(a^{\dagger}a\rho+\rho a^{\dagger}a-2a\rho a^{\dagger}\right)
-
\frac{\nu}{2}
\left(aa^{\dagger}\rho+\rho aa^{\dagger}-2a^{\dagger}\rho{a}\right),
\end{equation}
where $a$ and $a^{\dagger}$ are the annihilation and creation operators of 
the system (for example, an electro--magnetic field mode in a cavity), and 
$\mu,\ \nu$ are some constants depending on the system (for example, 
a damping rate of the cavity mode).

Since this is one of fundamental equations in quantum theory it is very 
important to construct the general solution. In \cite{BP} or \cite{WS} 
some methods to construct a solution are presented. However, a direct and 
clear method deriving a general solution has not been given as far as we know. 
It may be difficult to treat because 
(\ref{eq:quantum damped harmonic oscillator}) is an operator equation. 
In fact, in \cite{WS}; section 18.3.1 it is stated that 
``it is extremely difficult to solve the master equation directly". 
However, we can solve the equation (\ref{eq:quantum damped harmonic oscillator}) 
completely in the {\bf operator algebra level}.

First we separate the equation (\ref{eq:quantum damped harmonic oscillator}) 
into two parts and consider the one coming from the Lindblad form
\begin{equation}
\label{eq:Lindblad form}
\frac{\partial}{\partial t}\rho
=-
\frac{\mu}{2}
\left(a^{\dagger}a\rho+\rho a^{\dagger}a-2a\rho a^{\dagger}\right)
-
\frac{\nu}{2}
\left(aa^{\dagger}\rho+\rho aa^{\dagger}-2a^{\dagger}\rho{a}\right).
\end{equation}
Interesting enough, we can solve this equation completely by use of some 
method in \cite{KF1}, \cite{KF2}.

Let us rewrite (\ref{eq:Lindblad form}) to be more convenient by use of 
the number operator $N\equiv a^{\dagger}a$ ($[a,a^{\dagger}]={\bf 1}$)
\begin{equation}
\label{eq:Lindblad form-2}
\frac{\partial}{\partial t}\rho
=
\mu a\rho a^{\dagger}+\nu a^{\dagger}\rho{a}
-\frac{\mu+\nu}{2}(N\rho+\rho N +\rho)+\frac{\mu-\nu}{2}\rho
\end{equation}
where we have used $aa^{\dagger}=N+{\bf 1}$. Note that\ $[N,a]=-a$,\ \
$[N,a^{\dagger}]=a^{\dagger}$.

In order to solve the equation we can use the method in \cite{KF4} once more. 
For that we review a matrix representation of $a$ and $a^{\dagger}$ on the 
usual Fock space 
\[
{\cal F}=\mbox{Vect}_{\fukuso}\{\ket{0},\ket{1},\ket{2},\ket{3},\cdots \};
\quad \ket{n}=\frac{(a^{\dagger})^{n}}{\sqrt{n!}}\ket{0}
\]
like
\begin{eqnarray}
\label{eq:creation-annihilation}
a&=&\mbox{e}^{i\theta}
\left(
\begin{array}{ccccc}
0 & 1 &          &          &        \\
  & 0 & \sqrt{2} &          &        \\
  &   & 0        & \sqrt{3} &        \\
  &   &          & 0        & \ddots \\
  &   &          &          & \ddots
\end{array}
\right)\equiv \mbox{e}^{i\theta}b,\
a^{\dagger}=\mbox{e}^{-i\theta}
\left(
\begin{array}{ccccc}
0 &          &          &        &        \\
1 & 0        &          &        &        \\
  & \sqrt{2} & 0        &        &        \\
  &          & \sqrt{3} & 0      &        \\
  &          &          & \ddots & \ddots
\end{array}
\right)\equiv \mbox{e}^{-i\theta}b^{\dagger} 
\\
N&=&a^{\dagger}a=b^{\dagger}b=
\left(
\begin{array}{ccccc}
0 &   &   &   &        \\
  & 1 &   &   &        \\
  &   & 2 &   &        \\
  &   &   & 3 &        \\
  &   &   &   & \ddots
\end{array}
\right)
\end{eqnarray}
where $\mbox{e}^{i\theta}$ is some phase. Note that of course 
$[b,b^{\dagger}]={\bf 1}$.

For a matrix $X=(x_{ij})\in M({\cal F})$ 
\[X=
\left(
\begin{array}{cccc}
x_{11} & x_{12} & x_{13} & \cdots  \\
x_{21} & x_{22} & x_{23} & \cdots  \\
x_{31} & x_{32} & x_{33} & \cdots  \\
\vdots & \vdots & \vdots & \ddots
\end{array}
\right)
\]
we correspond to the vector $\widehat{X}\in 
{{\cal F}}^{\mbox{dim}_{\fukuso}{\cal F}}$ as
\begin{equation}
\label{eq:correspondence}
X=(x_{ij})\ \longrightarrow\ 
\widehat{X}=(x_{11},x_{12},x_{13},\cdots;x_{21},x_{22},x_{23},\cdots;
x_{31},x_{32},x_{33},\cdots;\cdots \cdots)^{T}
\end{equation}
where $T$ means the transpose. The following formula
\begin{equation}
\label{eq:well--known formula}
\widehat{AXB}=(A\otimes B^{T})\widehat{X}
\end{equation}
holds for $A,B,X\in M({\cal F})$.

Then (\ref{eq:Lindblad form-2}) becomes
\begin{eqnarray}
\label{eq:rho-equation}
\frac{\partial}{\partial t}\widehat{\rho}
&=&
\left\{
\mu a\otimes (a^{\dagger})^{T}+\nu a^{\dagger}\otimes a^{T}
-\frac{\mu+\nu}{2}(N\otimes {\bf 1}+{\bf 1}\otimes N+{\bf 1}\otimes {\bf 1})
+\frac{\mu-\nu}{2}{\bf 1}\otimes {\bf 1}
\right\}
\widehat{\rho} \nonumber \\
&=&
\left\{
\mu b\otimes b+\nu b^{\dagger}\otimes b^{\dagger}
-\frac{\mu+\nu}{2}(N\otimes {\bf 1}+{\bf 1}\otimes N+{\bf 1}\otimes {\bf 1})
+\frac{\mu-\nu}{2}{\bf 1}\otimes {\bf 1}
\right\}
\widehat{\rho},
\end{eqnarray}
so that the solution is formally given by
\begin{equation}
\label{eq:formal-solution}
\widehat{\rho}(t)=
\mbox{e}^{\frac{\mu-\nu}{2}t}
\mbox{e}^
{t
\left\{
\mu b\otimes b+\nu b^{\dagger}\otimes b^{\dagger}
-\frac{\mu+\nu}{2}(N\otimes {\bf 1}+{\bf 1}\otimes N+{\bf 1}\otimes {\bf 1})
\right\}
}
\widehat{\rho}(0).
\end{equation}

By the way, from the old lesson in \cite{KF1} we know a method to calculate 
(\ref{eq:formal-solution}) explicitly. 
By setting
\begin{equation}
\label{eq:K-generators}
K_{3}=\frac{1}{2}(N\otimes {\bf 1}+{\bf 1}\otimes N+{\bf 1}\otimes {\bf 1}),
\quad
K_{+}=b^{\dagger}\otimes b^{\dagger},\quad
K_{-}=b\otimes b,
\end{equation}
we can show the relations
\[
[K_{3},K_{+}]=K_{+},\quad [K_{3},K_{-}]=-K_{-},\quad 
[K_{+},K_{-}]=-2K_{3}
\]
easily. Namely we have the $su(1,1)$ algebra. The equation 
(\ref{eq:formal-solution}) can be written simply as
\begin{equation}
\label{eq:formal-solution-2}
\widehat{\rho}(t)=\mbox{e}^{\frac{\mu-\nu}{2}t}
\mbox{e}^{t\{\nu K_{+}+\mu K_{-}-(\mu+\nu)K_{3}\}}\widehat{\rho}(0),
\end{equation}
so we have only to calculate the term
\begin{equation}
\label{eq:exponential}
\mbox{e}^{t\{\nu K_{+}+\mu K_{-}-(\mu+\nu)K_{3}\}}.
\end{equation}

Now the disentangling formula in \cite{KF1} is helpful in calculating 
(\ref{eq:exponential}). 
We assume that there exists a unitary representation (group homomorphism) 
$\rho : SU(1,1) \subset SL(2;\fukuso)\ \longrightarrow\ U({\cal H})$ where 
${\cal H}$ is some Hilbert space related to the Fock space ${\cal F}$ above, 
and
\[
d\rho(k_{+})=K_{+},\quad d\rho(k_{-})=K_{-},\quad d\rho(k_{3})=K_{3}
\]
where $\{k_{+},k_{-},k_{3}\}$ are the generators of $su(1,1)$ algebra
\begin{equation}
\label{eq:generators}
 k_{+} = \left(
        \begin{array}{cc}
               0 & 1 \\
               0 & 0 
        \end{array}
       \right),
 \quad 
 k_{-} = \left(
        \begin{array}{cc}
               0 & 0 \\
              -1 & 0 
        \end{array}
       \right),
 \quad 
 k_{3} = \frac12 
       \left(
        \begin{array}{cc}
               1 &  0 \\
               0 & -1 
        \end{array}
       \right);
 \quad
 k_{-}\ne k_{+}^{\dagger}.
\end{equation}
It is very easy to check the relarions
\[
[k_{3},k_{+}]=k_{+},\quad [k_{3},k_{-}]=-k_{-},\quad 
[k_{+},k_{-}]=-2k_{3}.
\]

From (\ref{eq:exponential})
\begin{eqnarray}
\label{correspondence}
\mbox{e}^{t\{\nu K_{+}+\mu K_{-}-(\mu+\nu)K_{3}\}}
&=&
\mbox{e}^{t\{\nu d\rho(k_{+})+\mu d\rho(k_{-})-(\mu+\nu)d\rho(k_{3})\}}
=
\mbox{e}^{d\rho(t(\nu k_{+}+\mu k_{-}-(\mu+\nu)k_{3}))}  \nonumber \\
&=&
\rho\left(\mbox{e}^{t(\nu k_{+}+\mu k_{-}-(\mu+\nu)k_{3})}\right)
\equiv 
\rho\left(\mbox{e}^{tA}\right)
\end{eqnarray}
where
\begin{eqnarray*}
\mbox{e}^{tA}
&=&
\exp{
\left\{t
  \left(
   \begin{array}{cc}
    -\frac{\mu+\nu}{2} & \nu \\
    -\mu & \frac{\mu+\nu}{2}
   \end{array}
  \right)
\right\}
} \\
&=&
\left(
 \begin{array}{cc}
  \cosh\left(\frac{\mu-\nu}{2}t\right)-\frac{\mu+\nu}{\mu-\nu}
  \sinh\left(\frac{\mu-\nu}{2}t\right) & 
    \frac{2\nu}{\mu-\nu}\sinh\left(\frac{\mu-\nu}{2}t\right)  \\
  -\frac{2\mu}{\mu-\nu}\sinh\left(\frac{\mu-\nu}{2}t\right) &
  \cosh\left(\frac{\mu-\nu}{2}t\right)+\frac{\mu+\nu}{\mu-\nu}
  \sinh\left(\frac{\mu-\nu}{2}t\right)
 \end{array}
\right).
\end{eqnarray*}
The Gauss decomposition formula
\[
  \left(
   \begin{array}{cc}
     a & b \\
     c & d \\
   \end{array}
  \right)
=
  \left(
   \begin{array}{cc}
     1 & \frac{b}{d} \\
     0 & 1
   \end{array}
  \right)
  \left(
   \begin{array}{cc}
     \frac{1}{d} & 0 \\
     0 & d
   \end{array}
  \right)
  \left(
   \begin{array}{cc}
     1 & 0            \\
     \frac{c}{d} & 1
   \end{array}
  \right);\quad ad-bc=1
\]
gives

\begin{eqnarray*}
\mbox{e}^{tA}
=
  &&\left(
   \begin{array}{cc}
     1 & \frac{\frac{2\nu}{\mu-\nu}\sinh\left(\frac{\mu-\nu}{2}t\right)}
     {\cosh\left(\frac{\mu-\nu}{2}t\right)+\frac{\mu+\nu}{\mu-\nu}
      \sinh\left(\frac{\mu-\nu}{2}t\right)} \\
     0 & 1
   \end{array}
  \right)\times     \\
  &&\left(
   \begin{array}{cc}
     \frac{1}{\cosh\left(\frac{\mu-\nu}{2}t\right)+\frac{\mu+\nu}{\mu-\nu}
              \sinh\left(\frac{\mu-\nu}{2}t\right)} & 0 \\
     0 & \cosh\left(\frac{\mu-\nu}{2}t\right)+\frac{\mu+\nu}{\mu-\nu}
         \sinh\left(\frac{\mu-\nu}{2}t\right)
   \end{array}
  \right)\times     \\
  &&\left(
   \begin{array}{cc}
     1 & 0            \\
     -\frac{\frac{2\mu}{\mu-\nu}\sinh\left(\frac{\mu-\nu}{2}t\right)}
     {\cosh\left(\frac{\mu-\nu}{2}t\right)+\frac{\mu+\nu}{\mu-\nu}
      \sinh\left(\frac{\mu-\nu}{2}t\right)} & 1
   \end{array}
  \right).
\end{eqnarray*}
Moreover
\begin{eqnarray*}
\mbox{e}^{tA}
=
&\exp&
  \left(
   \begin{array}{cc}
     0 & \frac{\frac{2\nu}{\mu-\nu}\sinh\left(\frac{\mu-\nu}{2}t\right)}
     {\cosh\left(\frac{\mu-\nu}{2}t\right)+\frac{\mu+\nu}{\mu-\nu}
      \sinh\left(\frac{\mu-\nu}{2}t\right)}  \\
     0 & 0
   \end{array}
  \right)\times   \\
&\exp&
  \left(
   \begin{array}{cc}
     -\log\left(\cosh\left(\frac{\mu-\nu}{2}t\right)+\frac{\mu+\nu}{\mu-\nu}
                \sinh\left(\frac{\mu-\nu}{2}t\right)\right) & 0 \\
     0 & \log\left(\cosh\left(\frac{\mu-\nu}{2}t\right)+\frac{\mu+\nu}{\mu-\nu}
         \sinh\left(\frac{\mu-\nu}{2}t\right)\right)
   \end{array}
  \right)\times   \\
&\exp&
  \left(
   \begin{array}{cc}
     0 & 0            \\
     -\frac{\frac{2\mu}{\mu-\nu}\sinh\left(\frac{\mu-\nu}{2}t\right)}
     {\cosh\left(\frac{\mu-\nu}{2}t\right)+\frac{\mu+\nu}{\mu-\nu}
      \sinh\left(\frac{\mu-\nu}{2}t\right)} & 0
   \end{array}
  \right)     \\
=
&\exp&
   \left(\frac{\frac{2\nu}{\mu-\nu}\sinh\left(\frac{\mu-\nu}{2}t\right)}
     {\cosh\left(\frac{\mu-\nu}{2}t\right)+\frac{\mu+\nu}{\mu-\nu}
      \sinh\left(\frac{\mu-\nu}{2}t\right)}k_{+}\right)\times  \\
&\exp&
    \left(-2\log\left(\cosh\left(\frac{\mu-\nu}{2}t\right)+
           \frac{\mu+\nu}{\mu-\nu}\sinh\left(\frac{\mu-\nu}{2}t\right)\right)
    k_{3}\right)\times  \\
&\exp&
     \left(\frac{\frac{2\mu}{\mu-\nu}\sinh\left(\frac{\mu-\nu}{2}t\right)}
     {\cosh\left(\frac{\mu-\nu}{2}t\right)+\frac{\mu+\nu}{\mu-\nu}
      \sinh\left(\frac{\mu-\nu}{2}t\right)}k_{-}\right)
\end{eqnarray*}
by (\ref{eq:generators}). Since $\rho$ is a group homomorphism 
($\rho(XY)=\rho(X)\rho(Y)$) and 
the formula $\rho\left(\mbox{e}^{Lk}\right)=\mbox{e}^{Ld\rho(k)}$ where 
$k=k_{+},k_{3},k_{-}$ (see (\ref{correspondence})) we have
\begin{eqnarray*}
\rho\left(\mbox{e}^{tA}\right)
=
&\exp&
   \left(\frac{\frac{2\nu}{\mu-\nu}\sinh\left(\frac{\mu-\nu}{2}t\right)}
     {\cosh\left(\frac{\mu-\nu}{2}t\right)+\frac{\mu+\nu}{\mu-\nu}
      \sinh\left(\frac{\mu-\nu}{2}t\right)}d\rho(k_{+})\right)\times  \\
&\exp&
    \left(-2\log\left(\cosh\left(\frac{\mu-\nu}{2}t\right)+
           \frac{\mu+\nu}{\mu-\nu}\sinh\left(\frac{\mu-\nu}{2}t\right)\right)
    d\rho(k_{3})\right)\times  \\
&\exp&
     \left(\frac{\frac{2\mu}{\mu-\nu}\sinh\left(\frac{\mu-\nu}{2}t\right)}
     {\cosh\left(\frac{\mu-\nu}{2}t\right)+\frac{\mu+\nu}{\mu-\nu}
      \sinh\left(\frac{\mu-\nu}{2}t\right)}d\rho(k_{-})\right).
\end{eqnarray*}
As a result we have the disentangling formula
\begin{eqnarray}
\mbox{e}^{t\{\nu K_{+}+\mu K_{-}-(\mu+\nu)K_{3}\}}
=
&\exp&
   \left(\frac{\frac{2\nu}{\mu-\nu}\sinh\left(\frac{\mu-\nu}{2}t\right)}
     {\cosh\left(\frac{\mu-\nu}{2}t\right)+\frac{\mu+\nu}{\mu-\nu}
      \sinh\left(\frac{\mu-\nu}{2}t\right)}K_{+}\right)\times  \nonumber \\
&\exp&
    \left(-2\log\left(\cosh\left(\frac{\mu-\nu}{2}t\right)+
           \frac{\mu+\nu}{\mu-\nu}\sinh\left(\frac{\mu-\nu}{2}t\right)\right)
    K_{3}\right)\times  \nonumber \\
&\exp&
     \left(\frac{\frac{2\mu}{\mu-\nu}\sinh\left(\frac{\mu-\nu}{2}t\right)}
     {\cosh\left(\frac{\mu-\nu}{2}t\right)+\frac{\mu+\nu}{\mu-\nu}
      \sinh\left(\frac{\mu-\nu}{2}t\right)}K_{-}\right).
\end{eqnarray}

In the following we set for simplicity
\begin{eqnarray}
E(t)&=&\frac{\frac{2\mu}{\mu-\nu}\sinh\left(\frac{\mu-\nu}{2}t\right)}
     {\cosh\left(\frac{\mu-\nu}{2}t\right)+\frac{\mu+\nu}{\mu-\nu}
      \sinh\left(\frac{\mu-\nu}{2}t\right)},\quad
G(t)=\frac{\frac{2\nu}{\mu-\nu}\sinh\left(\frac{\mu-\nu}{2}t\right)}
     {\cosh\left(\frac{\mu-\nu}{2}t\right)+\frac{\mu+\nu}{\mu-\nu}
      \sinh\left(\frac{\mu-\nu}{2}t\right)}   \nonumber \\
F(t)&=&\cosh\left(\frac{\mu-\nu}{2}t\right)+
     \frac{\mu+\nu}{\mu-\nu}\sinh\left(\frac{\mu-\nu}{2}t\right).
\end{eqnarray}
Therefore (\ref{eq:formal-solution}) becomes
\[
\widehat{\rho}(t)=
\mbox{e}^{\frac{\mu-\nu}{2}t}
\exp\left(G(t)b^{\dagger}\otimes b^{\dagger}\right)
\exp\left(-\log(F(t))
    (N\otimes {\bf 1}+{\bf 1}\otimes N+{\bf 1}\otimes {\bf 1})\right)
\exp\left(E(t)b\otimes b\right)
\widehat{\rho}(0)
\]
under (\ref{eq:K-generators}). By use of $a=\mbox{e}^{i\theta}b$ and 
$a^{\dagger}=\mbox{e}^{-i\theta}b^{\dagger}$ in 
(\ref{eq:creation-annihilation}), and by some calculation
\begin{eqnarray}
\label{eq:approximate-solution-4}
\widehat{\rho}(t)
=
&&\frac{\mbox{e}^{\frac{\mu-\nu}{2}t}}{F(t)}
\exp\left(G(t)a^{\dagger}\otimes a^{T}\right)
\left\{
\exp\left(-\log(F(t))N\right)\otimes 
\exp\left(-\log(F(t))N\right)^{T}
\right\}\times    \nonumber \\
&&\qquad \ \ \exp\left(E(t)a\otimes (a^{\dagger})^{T}\right)
\widehat{\rho}(0)
\end{eqnarray}
where we have used $N^{T}=N$. Coming back to matrix form by use of 
(\ref{eq:well--known formula}) we obtain
\begin{eqnarray}
\rho(t)=
\frac{\mbox{e}^{\frac{\mu-\nu}{2}t}}{F(t)}
&&\sum_{n=0}^{\infty}
\frac{G(t)^{n}}{n!}(a^{\dagger})^{n}
[
\exp\left(-\log(F(t))N\right)\times
\nonumber \\
&&
\left.
\left\{
\sum_{m=0}^{\infty}
\frac{E(t)^{m}}{m!}a^{m}\rho(0)(a^{\dagger})^{m}
\right\}
\right.
\exp\left(-\log(F(t))N\right)
]
a^{n}.
\end{eqnarray}
This is indeed complicated !

\vspace{5mm}
Next we treat the full equation (\ref{eq:quantum damped harmonic oscillator})
\[
\frac{\partial}{\partial t}\rho=-i\omega(a^{\dagger}a\rho-\rho a^{\dagger}a)
-
\frac{\mu}{2}
\left(a^{\dagger}a\rho+\rho a^{\dagger}a-2a\rho a^{\dagger}\right)
-
\frac{\nu}{2}
\left(aa^{\dagger}\rho+\rho aa^{\dagger}-2a^{\dagger}\rho{a}\right).
\]
This can be rewritten as
\begin{equation}
\label{eq:full-equation}
\frac{\partial}{\partial t}\widehat{\rho}
=
\left\{
-i\omega K_{0}+\nu K_{+}+\mu K_{-}-(\mu+\nu)K_{3}
+\frac{\mu-\nu}{2}{\bf 1}\otimes {\bf 1}
\right\}
\widehat{\rho}
\end{equation}
in terms of $K_{0}=N\otimes {\bf 1}-{\bf 1}\otimes N$. 
Then it is easy to see
\begin{equation}
[K_{0},K_{+}]=[K_{0},K_{3}]=[K_{0},K_{-}]=0
\end{equation}
from (\ref{eq:K-generators}). That is, operators arising from 
the Hamiltonian of harmonic oscillator and its Lindblad form 
commute. Therefore 
\begin{eqnarray*}
\widehat{\rho}(t)
&=&
e^{-i\omega t K_{0}}
e^{t\{\nu K_{+}+\mu K_{-}-(\mu+\nu) K_{3}
+\frac{\mu-\nu}{2}{\bf 1}\otimes {\bf 1}\}}
\widehat{\rho}(0) \\
&=&
\mbox{e}^{\frac{\mu-\nu}{2}t}
\exp\left(-i\omega t K_{0}\right)
\exp\left(G(t)K_{+}\right)
\exp\left(-2\log(F(t))K_{3}\right)
\exp\left(E(t)K_{-}\right)
\widehat{\rho}(0) \\
&=&
\mbox{e}^{\frac{\mu-\nu}{2}t}
\exp\left(G(t)K_{+}\right)
\exp\left(\{-i\omega t K_{0}-2\log(F(t))K_{3}\}\right)
\exp\left(E(t)K_{-}\right)
\widehat{\rho}(0),
\end{eqnarray*}
so the general solution of the equation that we are looking for 
is just given by
\begin{eqnarray}
\label{eq:final form}
\rho(t)=
\frac{\mbox{e}^{\frac{\mu-\nu}{2}t}}{F(t)}
&&\sum_{n=0}^{\infty}
\frac{G(t)^{n}}{n!}(a^{\dagger})^{n}
\{
\exp\left(\{-i\omega t-\log(F(t))\}N\right)\times  \nonumber \\
&&
\left\{
\sum_{m=0}^{\infty}
\frac{E(t)^{m}}{m!}a^{m}\rho(0)(a^{\dagger})^{m}
\right\}
\exp\left(\{i\omega t-\log(F(t))\}N\right)
\}
a^{n}
\end{eqnarray}
by use of (\ref{eq:approximate-solution-4}).

If $\nu=0$, we have
\begin{equation}
\label{eq:corollary}
\rho(t)=
\mbox{e}^{-\left(\frac{\mu}{2}+i\omega\right)tN}
\left\{
\sum_{m=0}^{\infty}
\frac{\left(1-{e}^{-\mu t}\right)^{m}}{m!}a^{m}\rho(0)(a^{\dagger})^{m}
\right\}
\mbox{e}^{-\left(\frac{\mu}{2}-i\omega\right)tN}.
\end{equation}

\vspace{5mm}
To construct the general solution to the equation 
(\ref{eq:quantum damped harmonic oscillator}) is very important in not only 
Physics but also Mathematics, and we have finished it.

\vspace{5mm}
In this paper we revisited the quantum damped harmonic oscillator and 
constructed the general solution in the operator algebra level. 

The model is very important to understand several phenomena related to 
quantum open systems, so the general solution is needed. 
We finished it.

On the other hand we are studying some related topics from a different point 
of view, see \cite{KF3} and \cite{SR}. However, we make no comment on them 
in the paper and will report some results in another paper.

Lastly, we conclude the paper by stating our motivation. We are studying 
a quantum computation (computer) based on Cavity QED (see \cite{FHKW1} 
and \cite{FHKW2}), so in order to construct a more realistic model 
of (robust) quantum computer we have to study severe problems coming 
from decoherence. See also \cite{Scala et al-1}, \cite{Scala et al-2} 
for example. This is our future task.

\vspace{10mm}
\begin{center}
 \begin{Large}
  \textbf{Appendix}
 \end{Large}
\end{center}

\vspace{5mm}
In this appendix we explain our method in terms of the classical 
damped harmonic oscillator, see for example \cite{KF3} and \cite{SR}.

The differential equation is given by
\begin{equation}
\label{eq:classical damped harmonic oscillator}
\ddot{x}+2\gamma \dot{x}+\omega^{2}x=0\quad (\gamma > 0)
\end{equation}
where $x=x(t),\ \dot{x}=dx/dt$ and the mass is set to 1 for simplicity. 
In the following we treat only the case $\omega > \gamma$ (the case 
$\omega=\gamma$ may be interesting). 
By setting $y=\dot{x}$ the equation is rewritten as
\begin{equation}
\label{eq:classical damped harmonic oscillator-2}
\frac{d}{dt}
\left(
  \begin{array}{c}
    x \\
    y
  \end{array}
\right)
=
\left(
  \begin{array}{cc}
    0           & 1         \\
    -\omega^{2} & -2\gamma 
  \end{array}
\right)
\left(
  \begin{array}{c}
    x \\
    y
  \end{array}
\right).
\end{equation}
Noting
\[
\left(
  \begin{array}{cc}
    0           & 1         \\
    -\omega^{2} & -2\gamma 
  \end{array}
\right)
=
\left(
  \begin{array}{cc}
    -\gamma & 0        \\
     0      & -\gamma 
  \end{array}
\right)
+
\left(
  \begin{array}{cc}
     \gamma     & 1        \\
    -\omega^{2} & -\gamma 
  \end{array}
\right)
=
-\gamma{\bf 1}_{2}+k_{+}+\omega^{2}k_{-}+2\gamma k_{3}
\]
(see (\ref{eq:generators})) the general solution is given by
\begin{eqnarray}
\label{eq:general solution}
&&\left(
  \begin{array}{c}
    x(t) \\
    y(t)
  \end{array}
\right)
=
e^{-\gamma t}
e^{t(k_{+}+\omega^{2}k_{-}+2\gamma k_{3})}
\left(
  \begin{array}{c}
    x(0) \\
    y(0)
  \end{array}
\right)=       \nonumber \\
&&
e^{-\gamma t}
\left(
  \begin{array}{cc}
    \cos (\sqrt{\omega^{2}-\gamma^{2}}t)+
    \frac{\gamma \sin (\sqrt{\omega^{2}-\gamma^{2}}t)}{\sqrt{\omega^{2}-\gamma^{2}}}
    & 
    \frac{\sin (\sqrt{\omega^{2}-\gamma^{2}}t)}{\sqrt{\omega^{2}-\gamma^{2}}} 
    \\
    -\frac{\omega^{2} \sin (\sqrt{\omega^{2}-\gamma^{2}}t)}{\sqrt{\omega^{2}-\gamma^{2}}}
    &
    \cos (\sqrt{\omega^{2}-\gamma^{2}}t)-
    \frac{\gamma \sin (\sqrt{\omega^{2}-\gamma^{2}}t)}{\sqrt{\omega^{2}-\gamma^{2}}}
  \end{array}
\right)
\left(
  \begin{array}{c}
    x(0) \\
    y(0)
  \end{array}
\right).
\end{eqnarray}
For the details of calculation see \cite{KF3}.



\begin{thebibliography}{99}
%
\bibitem{BP}H. -P. Breuer and F. Petruccione : 
\newblock The theory of open quantum systems, 
\newblock Oxford University Press, New York, 2002.
%
\bibitem{WS}W. P. Schleich : 
\newblock Quantum Optics in Phase Space,
\newblock WILEY--VCH, Berlin, 2001.
%
\bibitem{KF1}K. Fujii : 
\newblock Introduction to Coherent States and Quantum Information Theory, 
\newblock quant-ph/0112090.
%
\bibitem{KF2}K. Fujii : 
\newblock Matrix Elements of Generalized Coherent Operators,
\newblock Yokohama Math. J, {\bf 53} (2007), 101,
\newblock quant-ph/0202081.
%
\bibitem{KF4}K. Fujii : 
\newblock An Approximate Solution of the Master Equation with the 
Dissipator being a Set of Projectors,
\newblock arXiv : 0708.4047 [quant-ph].
%
\bibitem{KF3}K. Fujii : 
\newblock Quantum Mechanics with Complex Time : A Comment to the Paper 
by Rajeev, 
\newblock quant-ph/0702148.
%
\bibitem{SR}S. G. Rajeev : 
\newblock Dissipative Mechanics Using Complex--Valued Hamiltonians, 
\newblock to appear in Annals of Physics, 
\newblock quant-ph/0701141.
%
\bibitem{FHKW1}K. Fujii, K. Higashida, R. Kato and Y. Wada : 
\newblock Cavity QED and Quantum Computation in the Weak Coupling Regime, 
\newblock J. Opt. B : Quantum and Semiclass. Opt, {\bf 6} (2004), 502, 
\newblock quant-ph/0407014. 
%
\bibitem{FHKW2}K. Fujii, K. Higashida, R. Kato and Y. Wada : 
\newblock Cavity QED and Quantum Computation in the Weak Coupling Regime II : 
Complete Construction of the Controlled--Controlled NOT Gate, 
\newblock Trends in Quantum Computing Research, Susan Shannon (Ed.), 
{\bf Chapter 8}, Nova Science Publishers, 2006 and 
Computer Science and Quantum Computing, James E. Stones (Ed.), 
{\bf Chapter 1}, Nova Science Publishers, 2007, 
\newblock quant-ph/0501046. 
%
\bibitem{Scala et al-1}M. Scala, B. Militello, A. Messina, S. Maniscalco, 
J. Piilo and K.-A. Suominen :
\newblock Cavity losses for the dissipative Jaynes-Cummings Hamiltonian 
beyond Rotating Wave Approximation, 
\newblock J. Phys. A: Math. Theor. {\bf 40} (2007), 14527,
\newblock arXiv:0709.1614 [quant-ph].
%
\bibitem{Scala et al-2}M. Scala, B. Militello, A. Messina, S. Maniscalco, 
J. Piilo and K.-A. Suominen :
\newblock Population trapping due to cavity losses,
\newblock arXiv:0710.3701 [quant-ph].
%
\end{thebibliography}
\end{document}